# The Digital Divide in Canada and the Role of LEO Satellites in Bridging the Gap

Tuheen Ahmmed[1,2], *Member, IEEE,* Afsoon Alidadi[1], *Member, IEEE,* Zichao Zhang[1], *Member, IEEE,* Aizaz U. Chaudhry[1], *Senior Member, IEEE,* and Halim Yanikomeroglu[1], *Fellow, IEEE*

[1]Department of Systems and Computer Engineering, Carleton University, Ottawa, ON K1S 5B6, Canada
[2]Ericsson Canada, Inc., Ottawa, ON K2K 2V6, Canada
Corresponding author: Tuheen Ahmmed (tuheen.ahmmed@ericsson.com)

***Abstract*—Overcoming the digital divide in rural and remote areas has always been a big challenge for Canada with its huge geographical area. In 2016, the Canadian Radio-television and Telecommunications Commission announced broadband Internet as a basic service available for all Canadians. However, approximately one million Canadians still don't have access to broadband services as of 2020. The COVID-19 pandemic has made the situation more challenging, as social, economic, and educational activities have increasingly been transferred online. The condition is more unfavorable for Indigenous communities. A key challenge in deploying rural and remote broadband Internet is to plan and implement high-capacity backbones, which are now available only in denser urban areas. For any Internet provider, it is almost impossible to make a viable business proposal in these areas. For example, the vast land of the Northwest Territories, Yukon, and Nunavut's diverse geographical features present obstacles for broadband infrastructure. In this paper, we investigate the digital divide in Canada with a focus on rural and remote areas. In so doing, we highlight two potential solutions using low Earth orbit (LEO) constellations to deliver broadband Internet in rural and remote areas to address the access inequality and the digital divide. The first solution involves integrating LEO constellations as a backbone for the existing 4G/5G telecommunications network. This solution uses satellites in a LEO constellation to provide a backhaul network connecting the 4G/5G access network to its core network. The 3rd Generation Partnership Project already specifies how to integrate LEO satellite networks into the 4G/5G network, and the Canadian satellite operator Telesat has already showcased this solution with one terrestrial operator, TIM Brasil, in their 4G network. In this way, users can seamlessly access broadband Internet via their mobile terminals. The second solution is based on the direct use of LEO constellations, such as Starlink, which are now operating in Canada, to deliver broadband Internet. As LEO satellites fly lower, their round-trip latency is lower, and the user terminals can receive Internet signals as long as they are pointing at the sky. An in-depth discussion of both solutions is presented in this work.**

## I. Introduction

THE Internet is one of the most important advances in human history. In the 21st century, it is like electricity or water: a lifeline for human communication, financial and economic systems, education and research, and a source of entertainment and well-being. During the COVID-19 pandemic, it has become obvious that broadband connectivity implies much more than social media. From employment to government and private services, from education to healthcare, these essential aspects of our lives depend on fast and reliable Internet connectivity.

A gap in Internet and technology access, known as the *digital divide*, has consistently increased due to the pandemic. Rural and remote communities in Canada, more specifically Indigenous communities, have been impacted the most. However, with the advent of telecommunication technologies, low population density and greater geographical distances should no longer be blamed for access inequality.

In most rural and remote areas in Canada, broadband Internet is either not available or its speed is not adequate for cloud-based software applications, video conferencing, online classes, and learning resources. Canada's current target is 50 Mbps download and 10 Mbps upload (50/10 Mbps) broadband Internet as a basic service for all Canadians, including rural households and businesses. The Government of Canada's venture aims for 90% of Canadians to have 50/10 Mbps broadband Internet by the end of 2021, for 95% to have this service by 2026, and for the remaining hardest-to-reach 5% regions to have this by 2030 [1]. To achieve this target, a constellation of small satellites in low Earth orbit (LEO) working together as a communications network presents an attractive solution for access inequality in Canada.

One of the main advantages of LEO satellites, as they orbit at an altitude of between 500 and 2,000 km, is that they can cover different types of terminals, like dedicated ground stations, 4G eNBs (evolved Node Bs), 5G gNBs (next generation Node Bs), satellite dish-type antennas, etc. The lower orbital altitude means lower latency for sensitive tasks, such as video calls and online gaming. However, the lower altitude also means that LEO satellites cover less of a footprint, and that at least 40 to 80 of them need to be deployed to provide uninterrupted services. Nevertheless, the cost of a LEO satellite is lower as it weighs less than a traditional geostationary satellite, normally less than 500 kg; LEO satellites can also be manufactured more quickly (less than one year); and up to 143 of them can be deployed via ridesharing with one rocket [2][3]. As there are great benefits of building LEO constellations, exploiting LEO satellites can be a better option to deliver higher data rates through broadband Internet and serve the unserved and underserved communities in Canada.

There are two ways LEO constellations can be used for faster deployment of broadband Internet in rural and remote areas. One way is to integrate LEO constellations with terrestrial infrastructure. Integrating LEO constellations with existing communication networks (such as 4G/5G) is a promising way of leveraging existing equipment to deploy broadband Internet in rural and remote areas. This concept is called backhauling. Its biggest advantage is that 4G/5G user equipment is fully supported, and no additional terrestrial infrastructure is required [2]. One example is the Canadian satellite operator Telesat, which is a pioneer in backhauling. Its constellation will be composed of 298 LEO satellites with a future plan to expand the constellation to 1,671 [4].

Another way that LEO satellites can be used for rural and remote broadband Internet access is through user terminals (UTs), which can communicate directly with satellites in LEO constellations. This is a more direct way of leveraging the ongoing deployment of mega-constellations that have a footprint in Canada; for example, SpaceX plans to deploy around 42,000 LEO satellites (12,000 authorized, and 30,000 requested for authorization) once in full service [5].

This paper provides a comprehensive overview of access inequality in Canada and an in-depth discussion of potential solutions to deliver broadband Internet via LEO satellite constellations to bridge the digital divide. More specifically, the contributions of the paper are the following:

- A discussion of access inequality and the digital divide in Canada with a focus on hard-to-reach areas and Indigenous communities.
- An overview of commercial LEO constellations that plan to operate in Canada, such as Telesat's Lightspeed and SpaceX's Starlink, including their technological details.
- A consideration of two potential solutions based on LEO satellites to deliver broadband Internet for bridging the digital divide in rural and remote areas in Canada.

## II. Related Research

Many proposals and pilot projects have already been attempted. However, they have not been very successful at providing adequate broadband Internet in hard-to-reach areas. One of the biggest projects to address underserved regions around the world was project Loon (parent company Alphabet Inc.). Starting in 2011, its mission was to provide Internet access in remote areas of the world. Loon's technology sent gas-filled balloons into the stratosphere. The onboard communications equipment sent Internet signals back down to the Earth and its mobile coverage was 200 times greater than a terrestrial cell tower. Though it showed successful use cases, for example, deploying balloons to the skies above Puerto Rico in 2017 after Hurricane Maria damaged the island's communications infrastructure, and providing cellular service by high-altitude balloons to Kenya's rural millions, it was shut down over cost concerns in 2021.

In Canada, three major optical fiber broadband projects are present with heavy provincial government subsidies as an alternative to private Internet Service Providers, such as Rogers and Bell. Two of these are in municipal areas: Calgary and Toronto. All of them are fiber expansions and take advantage of the existing dark fiber optic backbone. Only Quebec's Villages Branchés (Connected Villages) subsidy program is for rural broadband expansion [6]. The problems of these fiber broadband initiatives are that they are expensive, time-consuming, and it is hard for them to compete against private networks. Furthermore, laying fiber cable in Canada requires permits which could potentially delay any project.

Some other initiatives have also aimed to deliver Internet to remote areas. However, OneWeb went bankrupt in March 2020 (due to the high cost of manufacturing and launching satellites) before being bailed out in November 2020 by the British government and others. Other companies, such as Amazon and SpaceX, have continued efforts to provide Internet connectivity in hard-to-reach places. The Canadian satellite operator Telesat is also in this race.

Any big aerospace project, such as project Loon, has two major challenges: a technological challenge in delivering Internet in rural and remote areas, and a business model challenge. While the technological challenge can be overcome through new innovations, the challenge for businesses is hard to overcome due to the astronomical cost of terrestrial infrastructure deployment in rural and remote areas.

## III. Access Inequality in Rural and Remote Areas in Canada

*Rural and remote* areas are often thought to be similar, but there are differences between them. Geographic areas that are located outside urban areas are known as rural areas. According to the Canadian Radio-television and Telecommunications Commission (CRTC), rural communities are defined as areas with a population of less than 1,000 or a density of 400 or fewer people per square kilometer. Statistics Canada describes a remote community as a settlement that is either at a long distance from larger settlements or lacks transportation links that are common in more populated areas. 19% of Canadians live in rural and remote communities, and the broadband Internet speed is inadequate in these areas.

The *digital divide in Canada* refers to the gap between Canadians with high-speed broadband Internet and those without. The COVID-19 pandemic has widened this gap as Canadians have increasingly shifted their daily activities online. Broadband Internet is tied to many factors which result in a digital divide. Children suffering from this are unable to submit their homework online or join online classes. The digitally disconnected lose out on professional and educational opportunities and have difficulty accessing healthcare as well as government and social services [6].

More specifically, the Northwest Territories, Yukon, and Nunavut, where a significant percentage of the population consists of Indigenous people, do not have access to high-speed broadband networks. Low population density, rough terrain, and coastal areas have always been barriers to the development of broadband infrastructure. Internet Service Providers do not have sufficient impetus to build broadband infrastructure in the North due to the significant upfront investment and low return. So, Indigenous communities have limited access to these services.

*Indigenous peoples* are the original peoples of North America and their descendants. More than 1.67 million people in Canada are self-identified as Indigenous people. They are the fastest-growing as well as the youngest people in Canada. The distribution of Indigenous populations across Canada is visualized in Figure 1. It includes the total percentage of Indigenous people in each province and territory in comparison with its total population [7].

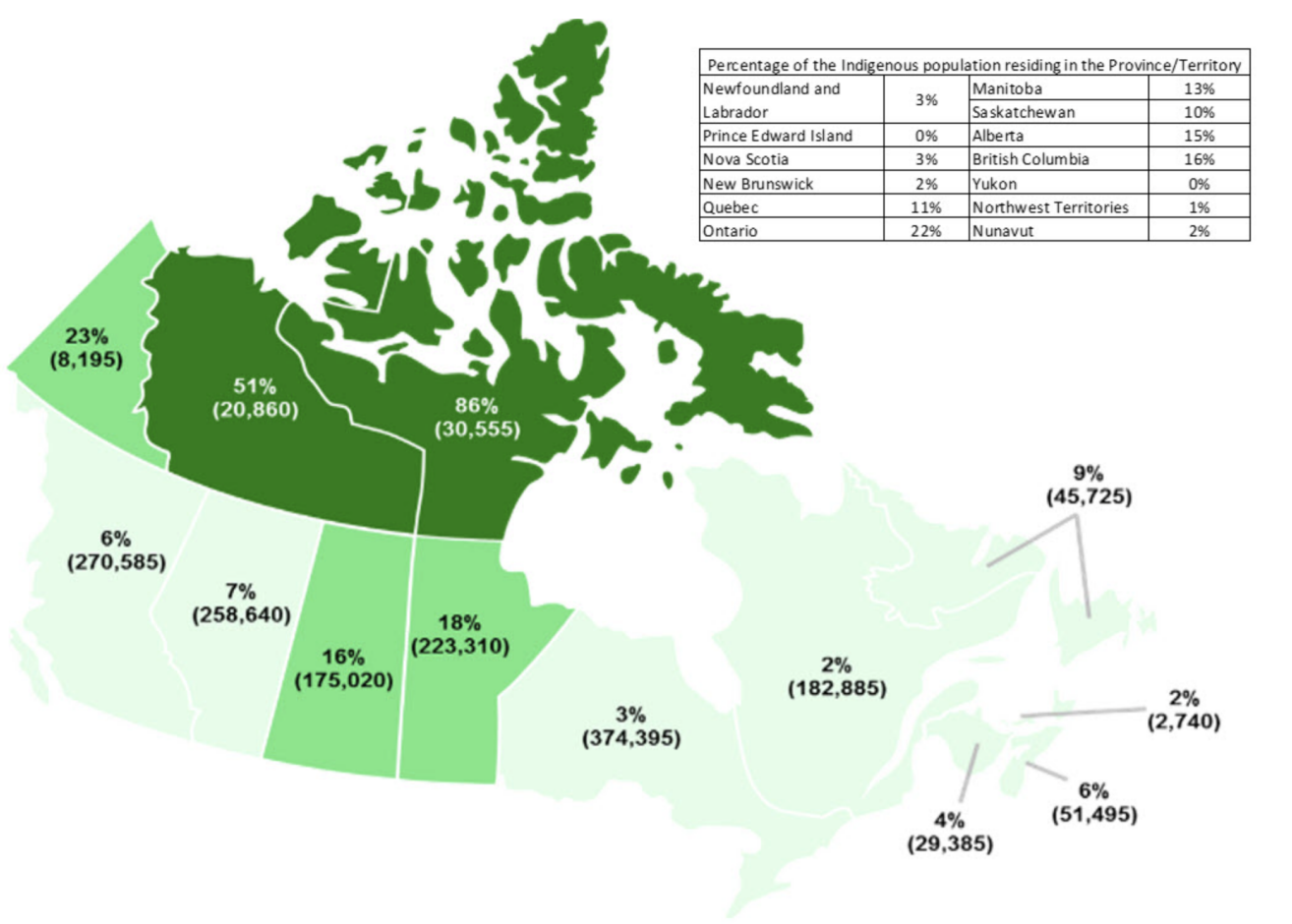

Figure 1: Distribution of Indigenous peoples across Canada [7].

Three groups of Indigenous peoples are recognized by the Canadian Constitution: First Nations, Inuit, and Métis. *First Nations* are divided into Registered Indians and Non-Status Indians. 49% of the Indigenous population are Registered Indians. *Inuits* predominantly reside in the Arctic and represent 4% of the total Indigenous population. Inuit Nunangat covers a third of Canada's landmass and 50% of its coastline; 73% of Inuits live there. Meanwhile, *Métis* are mixed European and Indigenous communities who live mainly in the three Prairie Provinces: Manitoba, Saskatchewan, and Alberta; 32% of the Indigenous population is Métis [7].

Every year, the CRTC publishes a detailed report about broadband connectivity. According to their Communications Monitoring Report published in 2020 based on 2019 data, only 45.6% of rural households have access to 50 Mbps download and 10 Mbps upload broadband services, compared with 98.6% of urban households.

The situation is more severe for First Nations reserves across Canada, where only 34.8% of First Nations people have a 50/10 Mbps connection as indicated in Figure 2. The target connectivity also varies significantly across provinces and territories, as we can see in the same figure. New Brunswick and British Columbia possess the highest speeds of Internet services, and 50 Mbps or faster speeds are available to 95.3% and 70.1% of the population, respectively. Only 5 Mbps download speed is available on First Nations reserves in Northwest Territories, Yukon, Newfoundland, and Labrador. Households in First Nations reserves in some other provinces—for example, Alberta, Manitoba, Ontario, and Saskatchewan—barely have 50/10 Mbps broadband services [8].

The economic divide is particularly wide in rural and remote areas, where people do not have adequate access to applications needed for education and telehealth purposes.

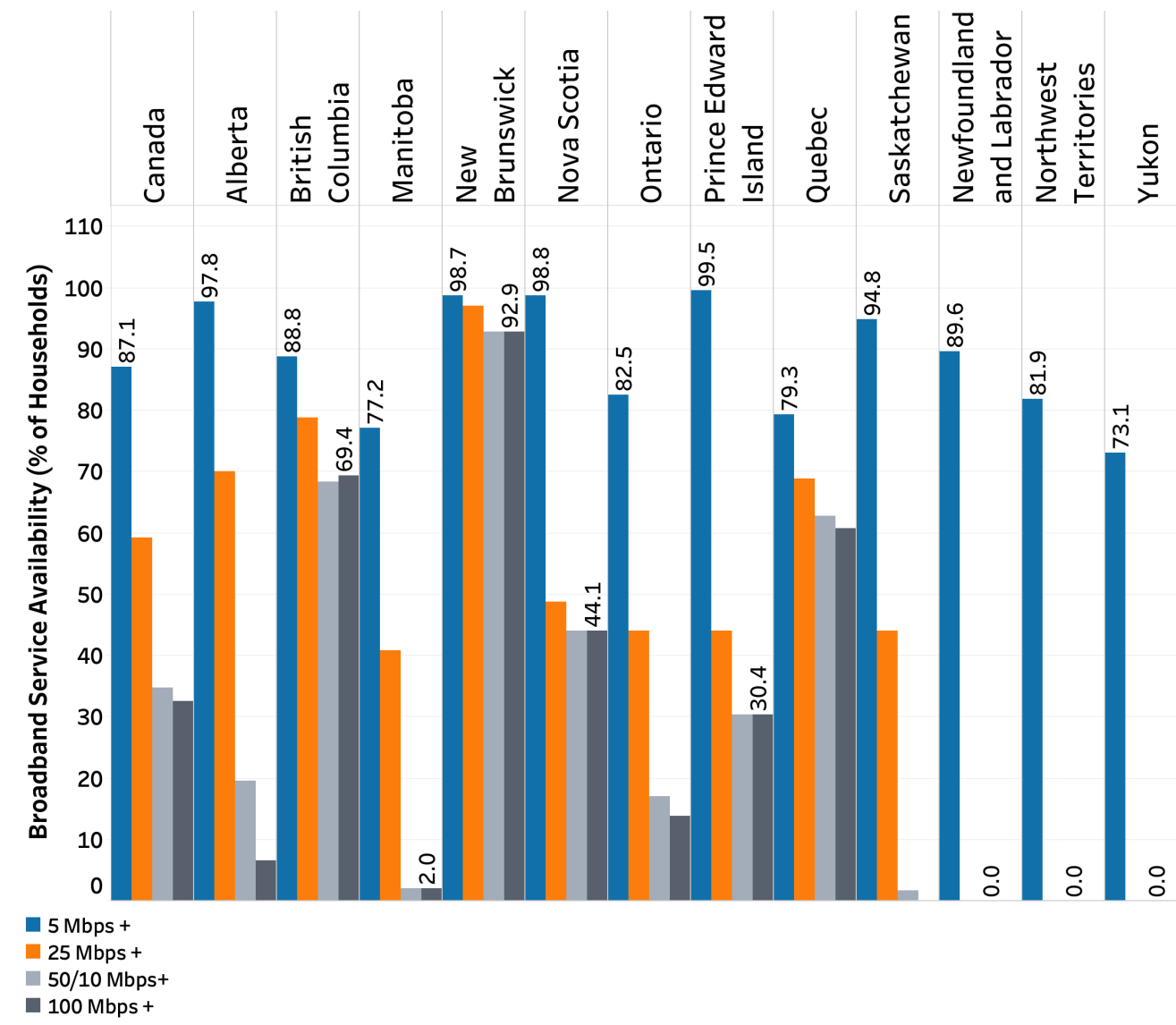

Figure 2: Availability of broadband services on First Nations reserves in 2019 in terms of speed across Canada (% of households).

Students and teachers are having a hard time connecting to online learning platforms, research tools, collaborative tools, and online classes. They are unable to engage in formal and informal education when in-person schools are closed, such as during the COVID-19 pandemic. Also, patients cannot access online healthcare facilities when in-person options are restrained or unsafe. With broadband connectivity, patients can leverage the benefits of telehealth applications and services, book appointments faster, consult expert doctors, have early diagnosis, eliminate travel time, and stay in known communities [9].

Stable and high-speed broadband connectivity is a must for economic development and better education and healthcare. The Truth and Reconciliation Commission of Canada recommends high speed broadband connectivity in Indigenous communities all over Canada to ensure education, healthcare, and economic growth for these communities. This would support the federal government's efforts for realizing Indigenous self-determination and reconciliation [9].

Canada is geographically enormous with a total area of 9.98 million square kilometers, but its population is only 38.65 million as of March 14, 2022. The high cost of terrestrial deployments makes the prospect of a lucrative business model in rural and remote areas unlikely. In this context, LEO satellites can be vital in delivering broadband Internet to the rural and remote areas of Canada to bridge the digital divide.

## IV. LEO Satellite Constellations in Canada

Having considerably wider coverage in comparison with other broadband access solutions, satellite constellations are being deployed around the globe to provide Internet access, especially in rural and remote areas. A satellite constellation is composed of a group of satellites, working together under central control and management, to provide a shared functionality for different users and entities.

One of the most important issues in satellite deployment is the launching cost. Until comparatively recently, the huge cost of satellite launching made it unaffordable for small corporations; only governmental agencies and large companies were capable of launching satellites into space. However, over the last two decades, innovative technologies and newer commercial rocket designs, which can accommodate multiple payloads, have reduced launch costs considerably leading to a significant increase in the number of satellite launches, especially in the private sector. This trend is expected to continue increasing over the next decade [10], as illustrated in Figure 3.

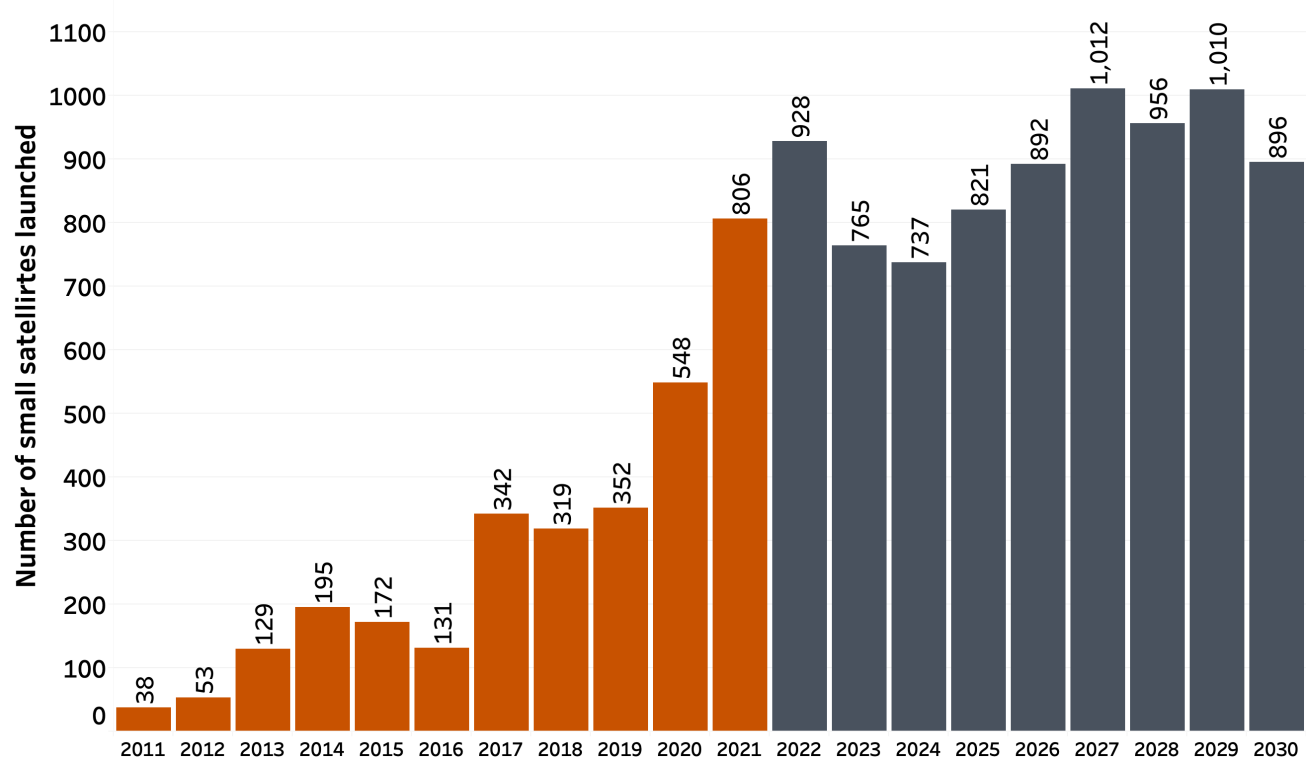

Figure 3: Launches of small satellites from 2011 to 2021, and projections from 2022 to 2030.

Satellite constellation systems are divided into three categories according to the distance between the satellite and the Earth's surface: geostationary Earth orbit (GEO), medium Earth orbit (MEO), and LEO [11]. Table I illustrates a comparison between the three categories of satellite constellations.

Considering the low latency and better frequency reuse features of LEO satellites, they have been of particular interest in providing fast Internet access to users, especially in hard-to-reach areas. In the following, we discuss two satellite constellations, which will be critical for enabling broadband Internet access in rural and remote communities in Canada to bridge the digital divide.

### A. *Telesat's Lightspeed*

Telesat started operating in 1969 as a commercial company and launched the first GEO domestic satellite, Anik A1, into space in 1972. In November 2016, Telesat proposed a ground-space integrated system named Lightspeed, which consists of 117 LEO satellites, to provide a capacity pool to serve Canadian Internet Service Providers and Mobile Operators. Later, this design was updated to a two-phased plan that added satellites to both polar and inclined orbits to improve performance accordingly [4]:

- Phase 1: 298 satellites; 6 polar orbits at an altitude of 1,015 km (13 satellites per orbital plane); and 20 inclined orbits at an altitude of 1,325 km (11 satellites per orbital plane).
- Phase 2: 1,671 satellites; 27 polar orbits at an altitude of 1,015 km (13 satellites per orbital plane); and 40 inclined orbits at an altitude of 1,325 km (33 satellites per orbital plane).

Telesat launched its first LEO satellite in 2018 as part of a modernized constellation for a new generation of global communications networking. The mixed orbital design of the Lightspeed constellation leads to wider coverage aligned with more signal routing choices, which will improve the connectivity of ground stations and end users.

The connection between Telesat's Lightspeed constellation and ground stations is through Ka-band (26.5–40.0 GHz) frequencies. This constellation utilizes the latest technologies in order to offer better capacity and coverage. In this regard, the Lightspeed satellites are equipped with data processing equipment, which provide full digital modulation, demodulation, and routing protocols, to improve capacity and flexibility. In addition, the laser inter-satellite links and phased array antennas used for communication with ground stations provide higher capacity and wider coverage in a more focused and systematic manner [4]. Lightspeed's highly advanced dynamic phased array antennas will be produced by Canadian space technology company MacDonald, Dettwiler & Associates, (MDA).

### B. *SpaceX's Starlink*

Starlink is an ongoing satellite constellation project by SpaceX with plans to deploy 42,000 LEO satellites (12,000 authorized, and 30,000 requested for authorization) to provide low latency, high data rate, and affordable Internet access worldwide. According to its FCC filings, this constellation has undergone several modifications. Its latest plan consisted of 4,408 LEO satellites operating in Ka and Ku bands [5]:

- 1,584 at a 550 km altitude and 53.0º inclination
- 1,584 at a 540 km altitude and 53.2º inclination
- 720 at a 570 km altitude and 70º inclination
- 348 at a 560 km altitude and 97.6º inclination
- 172 at a 560 km altitude and 97.6º inclination

Starlink's low altitude improves user experience from the perspective of latency. In addition, it facilitates satellite upgrades by reducing the time required to remove old satellites from orbit and send new ones back, leading to beneficial impacts upon the orbital debris mitigation.

TABLE I: Satellite Constellations Types

| Constellation Type | Altitude Range (km) | Round-Trip Latency (ms) | Number of Satellites for Global Coverage | Cost Per Satellite (million USD) | System Complexity | Orbital Period (hours) | Propagation Loss | Handover Requirement |
|---|---|---|---|---|---|---|---|---|
| GEO | ~36,800 | 400–600 | 3 | 100–400 | Low | 24 | High | Almost zero |
| MEO | 2,000–20,000 | 125–250 | 5–30 | 80–100 | Medium | 2–24 | Medium | Low |
| LEO | 500–2,000 | 30–50 | 40–80 | 0.5–45 | High | 1.5–2 | Low | High |

The Starlink network consists of ground stations, which are deployed across the country and are connected to Starlink's Points of Presence (POP) via fiber. Connection to these ground stations is established through two Ka-band parabolic antennas installed on the satellites. Starlink customers can connect to satellites directly by using a phased-array terminal that connects to a satellite's Ku-band RF beam. One of the main challenges in LEO constellations is the frequent handovers required to maintain user connectivity through the network. SpaceX's Starlink has made the handovers as smooth as possible through their phased-array technology at UTs and satellites, allowing both the satellite and user terminal antennas to adjust the direction of their RF beams.

## V. LEO Use Cases

There are two ways that LEO constellations can enable broadband Internet services. First, LEO constellations can be integrated as a backhaul in terrestrial networks. Usually, backhaul is implemented by optical fiber, fixed wireless, or cable; however, due to geography and cost, these are suboptimal solutions. To take advantage of existing telecommunications infrastructure, LEO satellites for backhaul between a terrestrial base station (BS) in a remote area and a core network (CN) is a promising solution. LEO satellites also enable backhauling between a moving BS (such as a BS on a train) and a CN. Both cases are illustrated in Figure 4. The 3rd Generation Partnership Project (3GPP) TS 38.821 (Release 16) specified in detail how to integrate LEO satellite networks with a 5G gNB. Also, it described corresponding physical layers, protocols, architecture, and radio resource management. The Central Unit and Digital Unit of an eNB/gNB are separable starting with Release 16 and are much easier to deploy and integrate with backhauling scenarios.

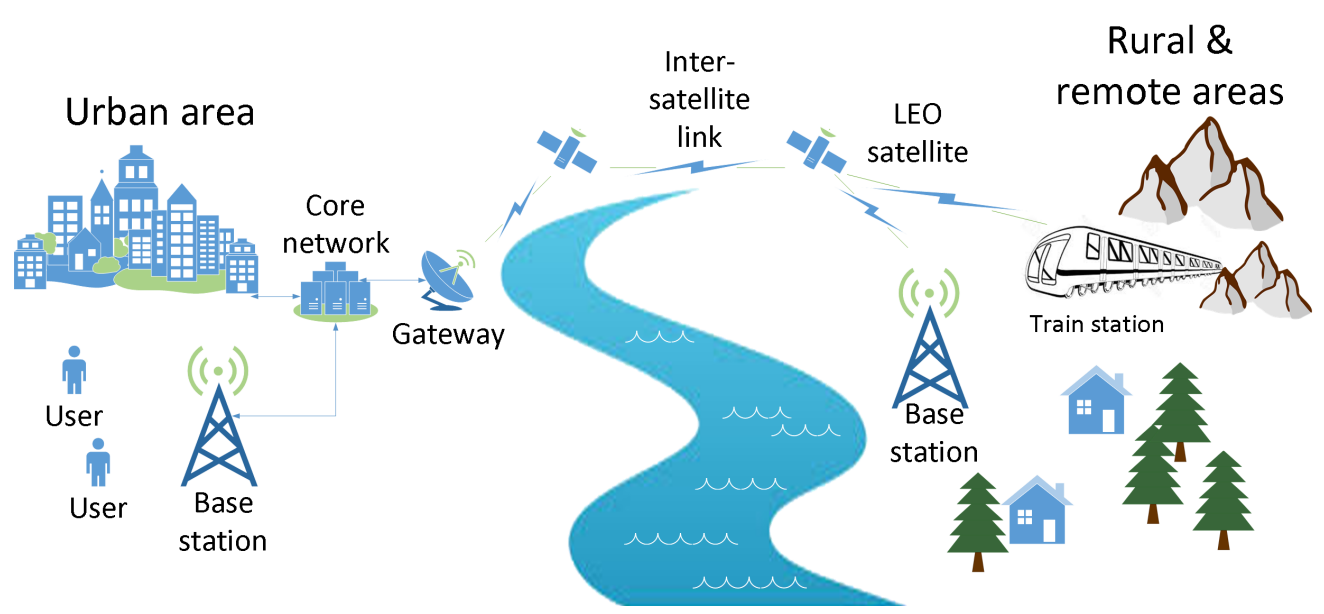

Figure 4: Use case showing Lightspeed's LEO satellites providing backhaul by integrating with existing infrastructure to support rural and remote communities.

Telesat's Lightspeed plans to offer a backhauling solution to enable broadband Internet access for growing communities, including rural and remote ones. One test case for Internet backhauling was demonstrated by Telesat and TIM Brasil, a leading telecommunications company in 4G coverage in Brazil. Telesat Lightspeed's Phase 1 LEO satellite was used to connect remote communities in Brazil to demonstrate mobile performance. One 85 cm Intellian parabolic antenna was used in uplink and downlink connections with this LEO satellite. 4G mobile data performance was measured and 38 ms latency was accomplished on average. All applications were tested properly and flawlessly. The applications tested were 1080p YouTube video streaming, video conferencing, WhatsApp voice over LTE, and interface compatibility [12].

The expected test results provide an incentive for operators like TIM Brasil to expand their services into rural and remote communities. Similar to Canada, Brazil has very good 4G coverage in urban areas. However, rural and remote communities cannot connect to the network in a cost effective way through fiber or additional cell towers because of significant distances and difficult terrain. These tests demonstrate how Telesat's Lightspeed can bring high-performance backhaul connectivity service to many underserved regions and bridge the Internet access inequality gap in Brazil [12]. OneWeb is in partnership with British Telecom (BT) to conduct similar tests. This involves testing the integration of OneWeb's LEO satellite technology with BT's existing terrestrial capabilities to develop a low-latency backhaul solution for bridging the digital divide across the UK.

As Telesat is from Canada, it will be easy to implement the same kind of Internet backhauling solutions with leading Canadian operators, such as Rogers, Bell, and Telus, to provide broadband Internet in rural and remote areas in Canada. Furthermore, the Government of Canada is investing \$1.44 billion in Telesat to deliver broadband Internet to the rural and remote communities. Telesat Lightspeed will enable broadband Internet in Canada starting from 2024, which will benefit Canadians, and especially Indigenous communities.

In the second category, UTs directly communicate with LEO satellites, as shown in Figure 5. As long as UTs consisting of satellite dish-type antennas are pointing towards the sky, end users can use high speed broadband Internet. With this solution, broadband access could even be provided for a small island in the middle of the ocean. Recently, SpaceX has set up a ground station on the Isle of Man, located in the Irish Sea between England and Ireland. The purpose is to provide broadband Internet to this remote area of England that cannot be reached by fiber broadband or 5G Internet. The 3GPP TS 38.811 (Release 15) defined these types of Non-Terrestrial Network (NTN) terminals. One good example in this category is Starlink that is operating in Canadian and other markets. Two other satellite companies, Amazon and OneWeb, are also working on direct broadband Internet for users worldwide, and Canadian users are expected to benefit as well. OneWeb is expected to launch broadband services in Canada in 2022.

Starlink UTs are small-sized outdoor dish antennas with a diameter of 58 cm (23 inches). In June 2021, SpaceX made an application to FCC to modify the UT size to 48 cm in diameter (approximately 18 inches). The UTs operate in the 14.0–14.5 GHz band for uplink and 10.7–12.7 GHz band for downlink, all in the Ku band. UTs can connect to SpaceX's Starlink satellites, which are visible on the horizon with a minimum 25 degrees elevation angle and a UT will be able to track Starlink's LEO satellites that are passing its visibility. To compensate for the loss in received power caused by changes in antenna gain and other factors, the phased array on UT is designed to maintain a constant received power level at the receiver [13].

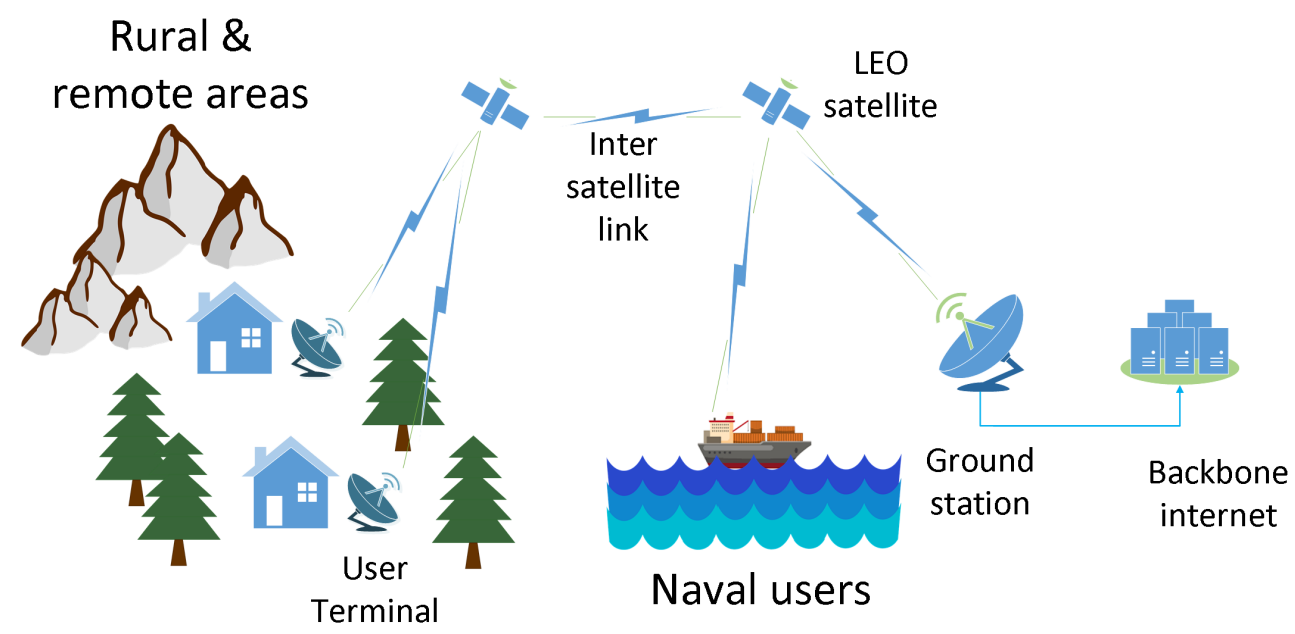

Figure 5: Use case illustrating Starlink's LEO satellites beaming Internet signals towards user terminals directly to provide broadband Internet in rural and remote areas.

A test result at the Makah Indian Reservations (a remote area of Washington State, USA) demonstrated the quality of Internet provided by Starlink. The downlink speed was 148 Mbps and the uplink speed was 13.6 Mbps in a household of four users that were simultaneously testing. Though the test speeds were lower than the USA average (180/66 Mbps), they were far better than the only current local Internet service, Centurylink (1/0.5 Mbps). In Canada, Starlink's downlink speed ranged from 53.61 Mbps in Ontario to 80.57 Mbps in Saskatchewan as per tests performed by Speedtest in Q1, 2021. This speed could potentially double when laser inter-satellite links are in place among Starlink satellites, and there will be fewer communications with ground stations and a lower number of ground stations will be required.

## VI. Risks and Concerns Associated with LEO Satellites

The satellite constellation industry is a trillion-dollar industry that has triggered the proliferation of small satellites. Aerospace companies are in a race to build mega-constellations in low Earth orbit. Apart from big companies such as SpaceX, Amazon, OneWeb, and Telesat, 200 other smaller companies are also planning to launch LEO satellites. When fully operational, SpaceX alone will have a mega-constellation of approximately 12,000 satellites and has plans to add another 30,000. As of December 2021, OneWeb has 394 in-orbit satellites, and they will deploy a total of 6,372 satellites. Amazon's Kuiper got approval from FCC for a deployment of 3,236 satellites.

For a LEO constellation, the total capital expenditure includes satellites launching cost, ground stations cost, the spectrum costs, and cost of integration with the existing terrestrial infrastructure. Due to ride sharing and multiple payloads, the launch cost has reduced from US$54,000/kg in 2000 to US$2,720/kg in 2018 [10]. In spite of this, it requires thousands of LEO satellites for a workable constellation. The lifetime of LEO satellites is about 5 years. Therefore, the satellites in a LEO constellation will need to be replaced, and a replacement cost of US$1–2 billion will be required periodically. Furthermore, an operational LEO constellation system costs around US$5–10 billion. So, significant upfront and ongoing investment is required, which makes the economic survivability of LEO constellations uncertain [14].

The success of a LEO constellation also depends on the affordability and reliability of its UTs. For example, for Canadian market, the UT cost from Starlink is CAD$699 and monthly service fee is CAD$129. Starlink is subsidizing the cost of its UTs to make them affordable but this could negatively impact their business case. Not all satellite constellation companies have deep pockets to survive in the long term without being profitable. Even though broadband connectivity via LEO constellations, such as Starlink, may exist in some Indigenous communities, people in these communities may not be able to afford the high monthly cost of the LEO satellites-based Internet or the high cost of its UTs.

New technologies such as content caching, network slicing, high gain antennas, and advanced satellite access protocol could play an important role in the success of the new LEO constellations. Content caching is a technique that is particularly helpful for video streaming. Unnecessary data transfer and delay is reduced by this technique by bringing the frequently requested content close to the users. However, direct-to-site business model based on UTs, such as Starlink, does not work the same way as content delivery networks do. So in a LEO constellation, every single view of a YouTube video by multiple users takes the full capacity of the satellite system and increases network load, and this may lead to congestion.

Canada has the highest spectrum license fees (average price per MHz of spectrum per person) among the Group of Seven countries, CAD$0.0351. The Canadian Government can implement the same approach like the governments in Mexico, New Zealand, and the USA have adopted, which involves setting aside dedicated spectrum for the Indigenous communities. Access to dedicated spectrum for Indigenous communities over their territories could be critical for connecting these communities to broadband Internet and for realizing Indigenous self-determination and reconciliation [9]. Another concern is interference between geostationary orbit and non-geostationary orbit (NGSO) satellite networks or between two different NGSO satellite networks as it can adversely affect their performance, for example, throughput degradation and received signal unavailability. This issue is highlighted in ITU-R's recommendation S.1419 and in the technical report entitled "A risk assessment framework for NGSO-NGSO interference" by the FCC Technological Advisory Council.

Managing and coordinating satellite traffic among companies is very challenging, and missteps can result in increased probabilities of collisions among LEO satellites. For example, in 2019, a Starlink satellite was about to collide with an Earth observation satellite. The rise of small satellites has the following two impacts as well: space debris and light pollution.

LEO satellites can break into debris upon collisions in the space environment. Then, collisions could take over low Earth orbit planes. A domino effect could produce more debris that would cause all satellites to become inoperable, a situation referred to as *Kessler Syndrome*. There are more than 12,000 trackable pieces of debris over 10 cm in diameter, whereas if we count sizes of 1 cm and up, there are about one million pieces of debris that are present in low Earth orbit [15].

*Light Pollution* is another major concern related to LEO satellites. LEO satellites create an array of bright lights which is visible in the clear night sky from many places all over the world. This puzzles astronomers as the lights emitted from satellites are brighter than 99% of everything else in Earth orbit.

## VII. Conclusion

Broadband Internet access will continue to shape people's lives. As the cost of deploying satellite constellations decreases, more hard-to-reach areas will have faster broadband connections, and the digital divide will be narrower. In this paper, we described the main opportunities of LEO small-satellite constellations that could dramatically improve broadband Internet connectivity in Canada as well as other rural and remote areas around the world. The two solutions we highlighted are practical and easy to implement for connecting rural and remote areas. The first solution, backhauling, leverages the existing telecommunications network. As backhauling can extend existing terrestrial networks to rural and remote areas using LEO satellites, people living in those areas will benefit. This will allow them to use broadband services over existing commercial smartphones. In the second solution, UTs can directly communicate with LEO satellites, and people in rural and remote areas would only need to purchase UTs to access broadband services. These solutions can be the fastest way to provide the target speed defined by the Canadian Government for all Canadians (i.e., 50 Mbps download and 10 Mbps upload) to bridge the digital divide.

## Acknowledgement

This work has been supported in part by the National Research Council Canada's (NRC) High Throughput Secure Networks program (CSTIP Grant #CH-HTSN-625) within the Optical Satellite Communications Consortium Canada (OSC) framework.

**Tuheen Ahmmed [M]** (tuheen.ahmmed@ericsson.com) is working as a Functional Systems Developer in Ericsson Canada Inc. He is pursuing MASc in Electrical and Computer Engineering at Carleton University. His research interests include 5G radio development and machine learning.

**Afsoon Alidadi [M]** (afsoonalidadishamsa@cmail.carleton.ca) is a Ph.D. student in the Department of Systems and Computer Engineering, Carleton University. Her research interests include the application of Artificial Intelligence and Machine Learning in HAPS and Satellite systems.

**Zichao Zhang [M]** (zichaozhang@cmail.carleton.ca) is a Ph.D. student in the Department of Systems and Computer Engineering, Carleton University. His research interest is faster than Nyquist signaling.

**Aizaz U. Chaudhry [SM]** (auhchaud@sce.carleton.ca) is a Senior Research Associate with the Department of Systems and Computer Engineering at Carleton University. His research interests include the application of machine learning and optimization in wireless networks. He is a licensed Professional Engineer in the Province of Ontario and a Senior Member of IEEE.

**Halim Yanikomeroglu [F]** (halim@sce.carleton.ca) is a Full Professor in the Department of Systems and Computer Engineering at Carleton University. His research group's current focus is aerial and satellite networks for the 6G era. His extensive collaboration with industry resulted in 39 granted patents. He is a Fellow of IEEE, and a Distinguished Speaker for both IEEE Communications Society and IEEE Vehicular Technology Society.